\documentclass{elsart}
\usepackage{graphicx}

\usepackage{amssymb}

\begin{document}
\begin{frontmatter}

\title{A Markovian Model Market - Akerlof's Lemmons and the Asymmetry of Information}

\author[ift]{Paulo F. C. Tilles},
\author[usp]{Fernando F. Ferreira},
\author[ift]{Gerson Francisco}
\author[usp]{Carlos de B. Pereira} and
\author[usp]{Flavia Mori Sarti},

\address[ift]{Instituto de F\'{\i}sica Te\'orica, Universidade Estadual
Paulista, R.  Pamplona 145, 01405-900 S\~ao Paulo, Brazil}

\address[usp]{Grupo Interdisciplinar de F\'{\i}sica da Informa\c{c}\~ao e Economia
 (GRIFE), Escola de Arte, Ci\^encias e Humanidades,
 Universidade de S\~ao Paulo,  Av. Arlindo Bettio 1000, 03828-000 S\~ao
 Paulo, Brazil}

\begin{abstract}
In this work we study an economic agent based model under different
asymmetric information degrees. This model is quite simple and can
be treated analytically since the buyers evaluate the quality of a
certain good taking into account only the quality of the last good
purchased plus her perceptive capacity $\beta $. As a consequence
the system evolves according to a stationary Markovian stochastic
process. The value of a product offered
    by the seller increases with quality according to the exponent $
\alpha $, which is a measure of technology. It incorporates all the
technological capacity of production
 systems such as education, scientific development and techniques that change the
 productivity growth. The technological level plays an important role to explain
  how the asymmetry of information may affect the market evolution in this model
  \cite{Chau}. We observe that, for high technological levels, the market can control
  adverse selection. The model allows us to compute the maximum asymmetric information
   degree before market collapse. Below this critical point the market evolves during a
    very limited time and then dies out completely. When $\beta $ is closer to 1
    (symmetric information), the market becomes more
profitable for high quality goods, although high and low quality
markets coexist. All the results we obtained from the model are
analytical and the maximum asymmetric information level is a
consequence of an ergodicity breakdown in the process of quality
evaluation.
\end{abstract}

\end{frontmatter}

\section{Introduction}

There is an increasing interest in microscopic models to investigate the
mechanism that drives the market \cite{Akerlof,Wang,Smith}. In particular,
researchers have focused on analytical and simple models that are able to
capture some real aspects of the economic activity.

Akerlof analyzed in \cite{Akerlof} the impact of asymmetric
information on the market. Agents trade in the market because they
share mutual benefits. However, when the information asymmetry
increases, high quality goods drop out from market and only low
quality goods can be traded, a situation called \textit{adverse
selection}.

Adverse selection is characterized as  pre-contractual opportunistic
behavior, since agents act in order to obtain more advantages, given
the market conditions and information available on the contracts
that regulate their commercial relations. Rules established in such
contracts end up attracting more intensely those agents that it
should repel (for example, the case of health insurance, which
attracts more sick than healthy individuals).

In the simple model proposed by Y-C Zhang \cite{Zhang}, Arkelof's
market failure is a special case. He revisited supply and demand
laws by setting quality and imperfect information as the key
ingredients. By varying continuously the information asymmetry
parameter one observes two regimes. For some intermediate level of
asymmetry consumers' benefits and firms' profits approach the
maximum. In this case, both cooperate and firms invest in educating
consumers to identify their goods' qualities. After this maximum
benefits still increase while profits decrease and consumers and
producers are in conflict. One solution for this dilemma is product
innovation.

Recently, Wang et. al \cite{Wang} introduced an agent based model to
address the impact of asymmetric information on the market
evolution. They showed the emergence of adverse selection from the
simulation of a very simple markovian model. For instance, the
market of used goods is not completely explained in terms of
asymmetric information \cite{Chau}. To improve market
understanding, it was introduced another variable called \textit{%
valuation ratio}, defined as the gap between buyers' valuation
(consumers' willingness to pay) and the sellers' valuation (firms'
cost) about the same good under the situation of complete
information. The coexistence of high and low quality is a
consequence of a trade-off between buyers' perception of product
quality and large valuation ratio.

It is worth to note that in the model we do not take into account
the quantity of produced goods. From the microeconomic theory, we
know that it is relevant to understand the supply and demand law.
However, as noted by Zhang \cite{Zhang}, products in the economy are
more complex. Both quality and information play important roles in
the understanding of how market behaves. The consumer's perception
depends on the amount of information available, becoming crucial for
the agent when it comes to evaluate the good and make decision.

We are not considering here the produced quantity. The relevant
aspect is only the product quality. Since in this study the amount of goods produced do not matter, the
model is suitable to describe the economy of information (software
companies, e-books, films production, and so for). In such context
the key feature are abstract aspects (knowledge, design and
creativity) and the replication cost is cheap (or the marginal cost
is constant). Consequently, the prototype is cheap to replicate but
hard to define the selling price, which depends on the product
quality  and the ability of consumers to recognize it.

In this work, we study a market model discussed above under some fixed
technological production levels and several asymmetric information degrees
\cite{Akerlof,Wang}. The model is developed on consumer's and firm's\
perceptions of value attributable to the good, regardless the market prices.

\section{Model}

There are $N_{s}$ sellers that offer goods with different integer quality
equally distributed in the range $[1,...,\kappa ]$ and $N_{b}$ buyers which
have access to all goods in the market. At each step the buyer $i$ chooses
one seller at random and evaluates the quality $q$ she is buying according
to the expression
\begin{equation}
q_{i}=\beta k+\left( 1-\beta \right) k_{a}^{i},
\end{equation}
where $\beta \in \lbrack 0,1]$ is the degree of asymmetric information, $k$
is the quality of goods informed by the seller to the consumers and $%
k_{a}^{i}$ is the quality of the last item purchased by the agent $i$. The
buyer recognizes the real quality of the item (the one attached by the
seller) only after purchasing it, and this quality is used as the next $%
k_{a}^{i}$. For $\beta =1$ the information is both perfect and
symmetric, that is, the two groups share the same quality
perception. As $\beta $\ decreases the consumers are no longer able
to precisely identify the products' quality (though they are not
aware of it). The asymmetry of information emerges from the
different perceptions between sellers and buyers about the same
product, and the buyers' inability to assign the product its true
quality raises an imperfect state of information (although it is
still perfect on the seller's side). The asymmetry reaches its
maximum when $\beta $ equals zero, making it impossible for the
buyers to have any idea about the quality of the good she is
purchasing.

Consumers and firms may disagree on how much a good is worth
(actually they do), so it is not expected they would agree on a
value assigned to it or even on how much should cost a difference in
the quality of a product. In this sense, we consider two distinct
ways for buyers and sellers to assign a value to the product,
depending on its quality. The consumers have a valuation function
\begin{equation}
V_{i}^{b}(k;k_{a}^{i})=Aq_{i},
\end{equation}
where $A$ is a rate of the buyer`s willingness to pay, while the sellers
compute the minimum value of their goods by taking into account the actual
manufacturing technological state of the economy $\alpha $ and the monetary
scale of sales $B$ via
\begin{equation}
V^{s}\left( k\right) =Bk^{\alpha },
\end{equation}%
with $\alpha >0$ and $B>0.$ A commercial transaction occurs only when
\begin{equation}
V^{b}(k;k_{a}^{i})\geq V^{s}(k),  \label{cond}
\end{equation}%
otherwise the buyer chooses randomly another seller in the next step.

The consumers' willingness to pay is linked to the value she
concedes to the good, which includes the valuation of a set of
attributes she values most at acquisition of a certain type of
product or service (for example, durability, quality, color, taste,
size, shape, sophistication, luxury, among others). If the sum of
those valuations given to the good falls below the firms production
cost (that is, in principle, the minimum value that the firm is
willing to accept to sell the good, which includes the firms'
opportunity costs), no trade transaction occurs. No consumer will
accept to pay a price higher than the value established by her
willingness to pay, since her perception of the attributes the good
will provide do not compensate the utility of the money she is
giving up.

When one computes the marginal value $\partial _{k}V^{s}(k)$, three
regimes may result according to the $\alpha $ parameter value. For
$\alpha <1$ the technological state of the economy is marginally
more efficient, i.e, cheaper than the case of $\alpha >1$ where the
derivative decreases as the quality increases (the marginal value is
constant when $\alpha =1)$. It is possible to associate $\alpha $
with the economy state of technology because a seller may produce
the same product with a lower cost by decreasing the $\alpha $\
value.

In a bargaining trading process the price will be somewhere between the
consumer`s willingness to pay $V^{b}\left( k;k_{a}\right) $ and the firm`s
cost $V^{s}\left( k\right) $. The discussion on the differences between
price and value involves lots of controversy, including the differentiation
between value in use and value in exchange,\ according to Adam Smith \cite%
{Smith}, "no trade occurs if the value in use of a good is lower
than its value in exchange". That is, whenever a trade occurs, the
value in use (utility of the good for the consumer) is usually
significantly higher than value in exchange (market price), and the
later should be at least equal to or higher than the production
cost. Based on this observation we define the relative profit as
$\gamma =\frac{V^{b}}{V^{s}}$ just to avoid introducing a
sophisticated price dynamics model.

Our focus is on the impact of asymmetric information under fixed
technological level in the evolution of the market. The buyer's quality
evaluation mechanism follows a markovian process since only the more recent
state is relevant for computing the next quality. We study the maximum
asymmetric information for the market to reach a stationary state. Also, we
discuss the consumer`s average observed quality for several asymmetric
information levels and the average number of transactions per round.

\section{The stationary probability distribution of the traded quality}

In each round a buyer chooses one seller at random, who will offer her a
product of quality $k$. The probability $p_{v}\left( k,k_{a}\right) $\ for
the transaction to occur is given by
\begin{equation}
p_{v}\left( k,k_{a}\right) =Pr\left[ V^{b}\left( k,k_{a}\right) \geq
V^{s}\left( k\right) \right] = Pr\left[ \beta k+\left( 1-\beta \right) k_{a}\geq \lambda {k}^{\alpha }%
\right] , \label{prob}
\end{equation}
where $\lambda =B/A$\ and $k_{a}$\ is the last product purchased by
the buyer. For this buyer's next round, her $k_{a}$\ will either be
the same or change to $k$, depending on the result of this
transaction. In this way one
can describe the buyer's behavior as a stochastic process, where her state $%
n $ is defined by the quality of her last purchased good, which may
remain the same if condition (\ref{cond})\ if not fulfilled (or if
she is trying to purchase the same quality) or undergo a transaction
$n\rightarrow m$\ whenever she tries to acquire a new good with
quality $m$. The transition between states must be probabilistic
since the new seller is chosen at random, and the use de Heaviside's
step function $\Theta \left( x\right) $\ guarantees that only
accessible states are reached. Describing the transition
probabilities as
\begin{eqnarray}
Pr\left( n\rightarrow m\right) =\frac{1}{N_{s}}\Theta \left[ \beta m+\left(
1-\beta \right) n-\lambda m^{\alpha }\right] , \\
Pr\left( n\rightarrow n\right) =\frac{1}{N_{s}}\Theta \left( n-\lambda
n^{\alpha }\right) + \frac{1}{N_{s}}\sum_{w=1}^{N_{s}}\Theta \left[ \lambda w^{\alpha }-\beta
w-\left( 1-\beta \right) n\right] .
\end{eqnarray}
one can define a transition matrix $T$ with elements $T_{nm}=Pr\left(
m\rightarrow n\right) $, and set a vector $\mathbf{P}_{t}$ describing the
probability of the buyer to be found in each state of the system. The
expression for the stochastic process given by $\mathbf{P}_{t+1}=T\mathbf{P}%
_{t}$, where the stationary solution $\mathbf{\Pi }=T\mathbf{\Pi }$\ may be
obtained by either diagonalizing the $T$\ matrix (in order to find the
eigenvector with eigenvalue equal to unity) or numerically by iterating the
equation until the convergence is reached.

The $\alpha $ parameter plays an important role in the model because it
affects the state of the economy depending on the values it may assume. By
considering a condition when a trade can not occur ($V^{b}<V^{s}$) and $%
\beta =1$, it is possible to find out the conditions for the market to
exist. If $\alpha \geq 1$, then we must have $\lambda \leq 1$, as for the
case when $\alpha <1$ the condition $\lambda \leq \kappa ^{1-\alpha }$ must
hold. Given that $\lambda $ satisfies either of this conditions, three
different market pictures arise, depending on the technological state $%
\alpha $: $\alpha >1$ gives rise to a forbbiden high quality region above $%
\hat{k}=\lambda ^{\frac{1}{1-\alpha }}$, independently of $\beta $; for $%
\alpha <1$ a forbbiden low quality region below $\hat{k}$ emerges when $%
\beta =1$, but it becomes allowed as $\beta $\ decreases; if $\alpha =1$ no
prohibitive region appears.

In Fig.1 we show the stationary solution $\Pi \left( k\right) $, with $%
\alpha =\frac{1}{2}$ and $\lambda =10$, for several asymmetric information
levels $\beta $. For perfect information ($\beta =1$) there is no
transaction below $k\leq \lambda ^{2}=100$. Thus, the distribution
probability $\Pi $ is uniform in the interval $\left[ \lambda ^{2},\dots
,\kappa \right] $, the buyers can purchase any goods available in the market
because there is no budget constraint in this model. When the asymmetry of
information arises, but is still close to 1, it becomes possible to purchase
products at a quality lower than $\lambda ^{2}$. For $\beta $ values much
lower than 1, the distribution probability $\Pi $ has a peak in this lowest
quality region indicating the adverse selection phenomena.
\begin{figure}[htb]
 \centering
 \fbox{\includegraphics[width=8cm, height=6cm]{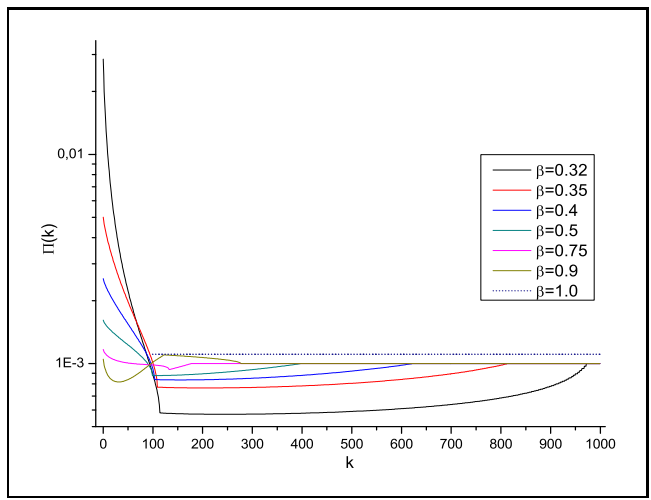}}
 \caption{\it Probability distribution $\Pi $ as a function of quality $k$. Information asymmetry rises the probability of
              negotiation for low quality}
 \label{fig:1}
\end{figure}
In the stationary state, the probability $\Pi (k)$ for a buyer to
be on the $k^{th}$ state may be associated with the probability of
his last purchased item be $k$, since $\mathbf{P}_{t+1}=\mathbf{P}_{t}=%
\mathbf{\Pi }$. This allows one to calculate the vendor's
probability to sell a product to one unspecified buyer using
(\ref{prob}),
\begin{equation}
p_{k}=\sum_{k^{\prime }=1}^{\kappa }p_{v}\left( k,k^{\prime }\right)
= \sum_{k^{\prime }=1}^{\kappa }\Theta \left[ \beta k+\left( 1-\beta \right)
k^{\prime }-\lambda k^{\alpha }\right] \Pi \left( k^{\prime }\right) . \label{pv}
\end{equation}
that is, the probability of the seller $k$ to perform a transaction given
that she was chosen by one buyer.

It is worth to note that, for fixed $\alpha <1$ there is a maximum
asymmetric information degree for the market to exist given by some
critical value $\beta _{c}$. This points out to the condition for
the occurrence of Akerlof`s lemmon market: association of low
quality goods with an environment of high information asymmetry
leads to low willingness to pay by consumers and high production
cost by firms ($\lambda >1$). For instance, taking $\lambda =10$ we
obtain $\beta _{c}=0.319$. In what follows we run all the simulation
for $\beta \in \left[ 0.32,\dots ,1\right] $. To get an
analytical curve for $\beta _{c}$ we observe that for a buyer in the state $%
k_{a}=1$ the trade condition below a certain asymmetric information degree $%
\beta _{c}$ is never satisfied, even if the buyer meets the highest
quality goods in the market (which she values the most), and the
market evolves to extinction. In other words, $k_{a}=1$ becomes an
absorbing state, and all agents (buyers) will eventually collapse to
it. Taking $V^{b}\left( \kappa ,1\right) =V^{s}\left( \kappa \right)
$ and $\beta =\beta _{c}$, we get
\begin{equation}
\beta _{c}\left( \kappa \right) =\frac{\lambda \kappa ^{\alpha }-1}{\kappa -1%
}.  \label{beta}
\end{equation}
Also if $V^{b}\left( \kappa ,1\right) =V^{s}\left( \kappa \right) $\ holds,
then for any state $n>1$ we have\ $V^{b}\left( \kappa ,n\right) \geq
V^{s}\left( \kappa \right) $, which means that if $k_{a}=1$\ is not an
absorbing state then there are no absorbing states in the system\footnote{%
There exist at least one state available for each $n>1$.}. In this
way, the maximum asymmetric information level is a consequence of an
ergodicity breakdown in the process of quality evaluation.

The completeness of the system (accessibility of all states) is
guaranteed if the condition
\begin{equation}
\kappa \geq \frac{\lambda -\beta }{1-\beta }
\end{equation}
holds, but it's only valid in the region $\beta <1$, since for
$\beta =1$ the existence of states where the buyer's valuation is
below the seller's level is not allowed.

From equation (\ref{beta}), one may see that there is an interval
where the parameter $\lambda $ has some interest. Noting that $0\leq
\beta \leq 1$ we get $\kappa ^{1-\alpha }\leq \lambda \leq \kappa
^{-\alpha }$. Beyond the lower limit of $\lambda $ the market is
never affected by the level of information since buyer may always
buy any product in the market. On the other extreme, the market
collapses because buyers underestimate all qualities, not paying the
value asked by the sellers.

\section{Statistics}

To obtain all statistical quantities of interest we need to take
into account the probability of a seller to be chosen by $n$ buyers
in one single run. Given $N_{s}$ sellers and $N_{b}$ buyers, let
$X_{k}$ be a random variable that represents the number of buyers
that choose the seller k in some round. So,
\begin{equation}
P\left( X_{n}=0\right) =\left( \frac{N_{s}-1}{N_{s}}\right) ^{N_{b}}.
\end{equation}
Each buyer chooses only one seller at each step. From the seller viewpoint
the probability to be chosen by exactly $n$ buyers is described by a
binomial distribution
\begin{equation}
P\left( X_{k}=n\right) =Pr\left( n\right) =\frac{N_{b}!}{n!\left(
N_{b}-n\right) !}\frac{\left( N_{s}-1\right) ^{N_{b}-n}}{N_{s}^{N_{b}}}.
\end{equation}

Now we obtain an expression for the probability of some seller $k$ to
perform exactly $j$ transactions, given that he was chosen by $n$ buyers ($%
n\geq j$). Again, since the probability of a seller $k$ actually sell to $%
Y_{k}=j$ buyers follows a binomial process where the Bernoulli trial has the
weights $p_{k}$ (\ref{pv}), the result is
\begin{equation}
P\left( Y_{k}=j|X_{k}=n\right) =\frac{n!}{j!\left( n-j\right) !}%
p_{k}^{j}\left( 1-p_{k}\right) ^{n-j}.  \label{pa}
\end{equation}%
Note that, when the system reaches the stationary state, we can use
the ergodic property to calculate the temporal average number of
transaction per run due to the seller $k$ as a sum over all states
of the system:
\begin{eqnarray}
\left\langle trans(k)\right\rangle & = \sum_{n=0}^{N_{b}}nP\left(
Y_{k}=n\right) = \sum_{n=0}^{N_{b}}n\sum_{i=n}^{N_{b}} Pr\left( i\right) P\left(
Y_{k}=j|X_{k}=n\right) \nonumber \\
& = \frac{N_{b}!}{N_{s}^{N_{b}}}\sum_{n=0}^{N_{b}}\sum_{i=n}^{N_{b}}\frac{%
\left( N_{s}-1\right) ^{N_{b}-i}p_{k}^{n}\left( 1-p_{k}\right) ^{i-n}}{%
\left( N_{b}-i\right) !\left( i-n\right) !\left( n-1\right) !}.
\end{eqnarray}
where $P\left( Y_{k}=n\right) $ is the probability of a seller sell exactly $%
n$ units of goods.

Let $P(k|k^{\prime })$ the probability that, given some buyer state $%
k^{\prime }$, some seller $k$ will do business. We write this as
\begin{equation}
P(k|k^{\prime })=\frac{\Theta \left[ \beta k+\left( 1-\beta \right)
k^{\prime }-\lambda k^{\alpha }\right] \Pi \left( k^{\prime }\right) }{%
\sum_{k^{\prime }=1}^{N_{b}}\Theta \left[ \beta k+\left( 1-\beta \right)
k^{\prime }-\lambda k^{\alpha }\right] \Pi \left( k^{\prime }\right) }
\end{equation}
So, the average of any quantity written as a function of $k$ and $k^{\prime
} $ denoted by $\hat{O}\left( k;k^{\prime }\right) $ can be obtained via
\begin{equation}
\left\langle O\left( k\right) \right\rangle =\sum_{k^{\prime }=1}^{N_{b}}%
\hat{O}\left( k;k^{\prime }\right) P(k|k^{\prime })  \label{aver}
\end{equation}

\section{Results and Discussions}

In this section we will analyze some quantities obtained from simulation to
check the results expressed in equation (\ref{aver}). Again, we set up $%
N_{s}=N_{b}=1000$, $\lambda =10$ and $\alpha =0.5$ since our
interest is focused on the analytical expression. Our main interests
are the individual averages of the number of transactions, the
valuation ratio and the observed quality by buyers. The average
number of transactions (per round) gives a measure of firms
efficiency to sell their products, but it must be analyzed
together with the valuation rate $\gamma \left( k\right) =V^{b}(k)/V^{s}(k)$%
, because companies in general tend to maximize their expected
profit according to a strategy that may  last longer than a single
run, i.e., a firm may choose either to make less transactions on a
time step with a higher profit or to sell her product to as many
consumers as possible with a lower margin. As far as the buyers'
perception play an important role in the model, each firm's observed
average quality tells us how the buyers evaluate the quality of its
product. If the consumer overestimates the quality of the product,
trade will occur only if the willingness to pay is high enough to
afford it, otherwise we expect the buyer to give up the purchase.
Quality underestimation may also occur on regions where the value
assigned by the consumer with perfect information is much higher
than the sellers, but it leads to a lower margin profit. So
asymmetric information is not always advantageous to firms, since
the average number of transactions and/or the relative gain may
decrease, and in high levels, it may even cause the market's
collapse.

When it comes to analyze the average number of transactions made by
each firm, the state of symmetric information forbids all sellers
with quality below the $k=100$ plateau to make any transaction,
because the consumers' evaluation is below the sellers' evaluation.
Above this plateau, each seller is
equally likely to sell her product, averaging one transaction each round.

As $\beta $\ decreases, one recognizes that a valley appears in the
region between high and low qualities. Even extremely low quality
products become tradeable, reaching almost the same transaction
level observed for that products with superior quality (Fig. 2).
\begin{figure}[htb]
 \centering
 \fbox{\includegraphics[width=8cm, height=6cm]{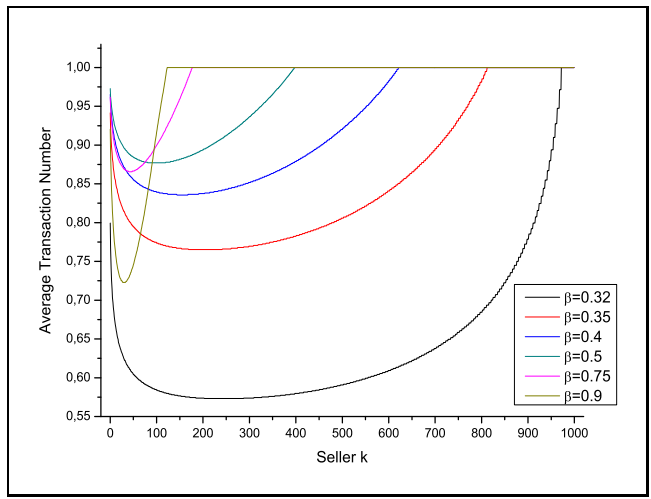}}
 \caption{\it Average Number of Transactions as a function of quality $k$.}
 \label{fig:2}
\end{figure}
The average valuation ratio $\left\langle \gamma \left( k\right)
\right\rangle$ (Fig. 3), in the symmetric information case, increase
 monotonically with the increasing of quality. By introducing
 asymmetry in the information low quality goods become more
 profitable. In other words, as $\beta $\ decreases to $\beta _{c}$
 the valuation ratio invert completely its behavior.
But it is in the quality mid-region $%
\left[ k\sim 50,k\sim 800\right] $ that an interesting behavior emerges. A
significant lower level of transactions appears, becoming larger as $\beta $%
\ approaches its critical level $\beta _{c}$ (affecting more
products during the transition), while the valley deepness appears
to decrease from $\beta \approx 0.9$\ to $\beta \approx 0.5$, and to
increase from $\beta \approx 0.5$ until it reaches its maximum level
at $\beta _{c}=0.32$, as one can see in Fig. 2.
\begin{figure}[htb]
 \centering
 \fbox{\includegraphics[width=8cm, height=6cm]{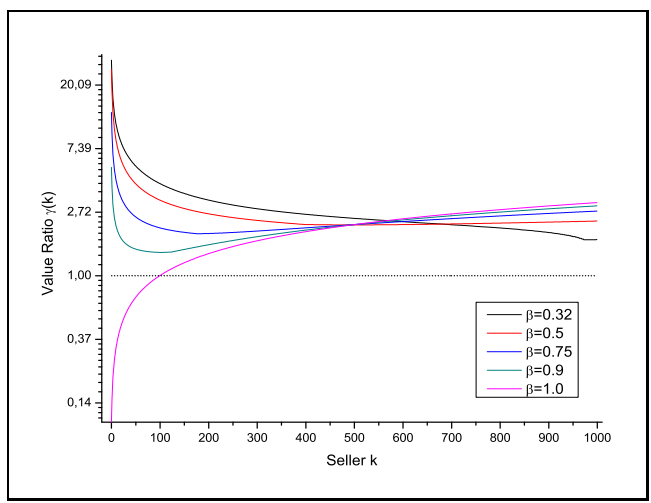}}
 \caption{\it Value ratio $\protect\gamma =V^{b}(k)/V^{s}(k)$, i.e. the gap between value given by buyer over
              the value due to the seller as a function of quality $k$.}
 \label{fig:3}
\end{figure}

The relative gain $\gamma \left( k\right) $ allows us to compare the
relative firms' profit as a function of the quality $k$ for several
$\beta $\ values. In the symmetric information case $\gamma $ is an
increasing function, showing that it is more profitable to produce
high quality goods.

Information asymmetry makes the low quality region become more
profitable, which can be interpreted as an incentive for firms
produce low quality goods. This is a consequence of the goods
quality overestimation by buyers. We note that the minimum profit
depends on $\beta $ and moves from low to high quality level, in
particular, for $\beta =\beta _{c}$ the minimum $\gamma $ is the
highest quality product.

It is important to note that when $\beta $ is close to $%
\beta _{c}$ it becomes difficult for buyer to recognize the real
product quality . Below the critical point $\beta =\beta _{c}$ the
buyers move away from the market because they become unable to
distinguish the goods' quality. This is the reason pointed out by
Arkelof to explain the market failure, caused by information
asymmetry associated with high variation in the quality of goods
produced.
\begin{figure}[htb]
 \centering
 \fbox{\includegraphics[width=8cm, height=6cm]{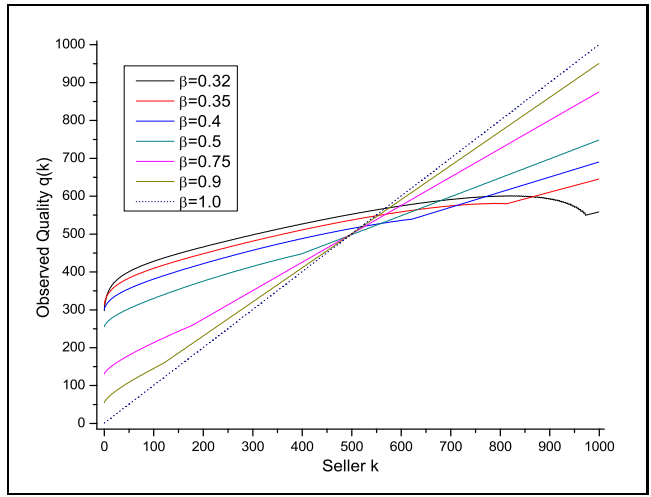}}
 \caption{\it Observed quality $q$ as a function of quality $k$.}
 \label{fig:4}
\end{figure}

The general model overview is that an environment of perfect and
symmetric information enables consumers to purchase products which
they think are fairly evaluated. When their perception $\beta $
decreases from $\beta =1.0$ to $\beta _{c}$, low quality goods are
traded in the market less frequently than those with high quality.
The effect of reducing the information level is the decrease in the
average rate of transactions. For a high information asymmetry
($\beta \approx \beta _{c}$), less transactions take place, mostly
concentrated on very low and very high quality level goods. When the
firms' quality decrease, the trade probability also decreases (see
Fig. 2) and so does the average transaction rate. However, we see in
the Fig. 1 that
there is a peak in the distribution $\Pi \left( k\right) $ in the region $%
k\leq \lambda ^{\frac{1}{1-\alpha }}$, which explains the increasing
transaction rate in this region for $\beta \approx \beta _{c}$.

\section{Conclusions}

The simplicity of the model allowed us to solve it analytically
using the Markovian chain approach. The existence of stationary
solutions is a condition for the systems to avoid the collapse. The
model can be split in three regimes, according to the value of
technological level $\alpha $. The role of the technology is to
increase productivity and to reduce cost. In particular, we focused
the analyzes on high technological regime ($\alpha =\frac{1}{2}$)
since the economy evolves toward increasing efficiency.

The role of the asymmetric information is to enhance the low quality
market. We showed that the consumers overestimate the quality of
goods below the medium quality level. Thus the relative profit
increases in the region of low quality, and a higher asymmetric
information will imply higher profits in the production of the low
quality goods. However, there is a critical asymmetric level above
which the market collapses. This is explained as the breakdown of
ergodicity, i.e., the asymmetry is so high that all those that buy
low quality goods fall down in a trap and stop to trade. From the
consumers' viewpoint, the incapacity to distinguish high and low
quality makes the agent leave the market.

From the seller' point of view, it is hard to decide which quality
would maximize her profit. It is necessary to know the value of
several parameters from the market. This is not the mission of
economics. In general Economics provides a theory that explains what
will happen given the market set up. The design of strategies is the
business of marketing science. The importance of general models is
provide tools that help us to address this issue.

\ack

FFF thanks the Conselho Nacional de Pesquisa (CNPq) for financial
support and Instituto de F\'{\i}sica Te\'orica for hospitality.

\end{document}